%% file: main.tex
\documentclass[letterpaper,twocolumn,10pt]{article}
\usepackage{usenix}

\usepackage[english]{babel}
\usepackage{tikz}
\usetikzlibrary{arrows.meta,positioning,fit,backgrounds,shapes.geometric}
\usepackage{pgfplots}
\pgfplotsset{compat=1.17}
\usepackage{xurl}
\usepackage{amsmath}
\usepackage{amssymb}
\usepackage{mathtools}
\usepackage{pifont}
\usepackage{algorithm}
\usepackage{algorithmicx}
\usepackage{algpseudocode}
\usepackage{amsmath}
\usepackage{amsthm}
\usepackage{enumitem}
\usepackage{microtype}
\usepackage{graphicx}
\usepackage{wasysym}
\usepackage{soul}

\usepackage{bbding}
\usepackage{subfigure}
\usepackage{booktabs}
\usepackage[table]{xcolor}
 \usepackage{balance}
 \usepackage{multirow}
\usepackage{multicol}
\usepackage{hyperref}
\usepackage[all]{nowidow}
\usepackage{color, colortbl}
\usepackage{xspace}

\definecolor{Gray}{gray}{0.9}

\usepackage{listings}
\usepackage{listingsutf8}

\lstdefinestyle{json}{
  basicstyle=\ttfamily\footnotesize,
  breaklines=true,
  xleftmargin=1.5em,
  numbers=none
}

\lstdefinestyle{python}{
  basicstyle=\ttfamily\footnotesize,
  breaklines=true,
  xleftmargin=1.5em,
  numbers=none,
  keywordstyle=\color{blue},
  stringstyle=\color{mGreen},
  commentstyle=\color{mGray}
}

\lstdefinestyle{cpp}{
  basicstyle=\ttfamily\footnotesize,
  breaklines=true,
  xleftmargin=1.5em,
  numbers=none,
  keywordstyle=\color{blue},
  stringstyle=\color{mGreen},
  commentstyle=\color{mGray}
}

\lstdefinestyle{c}{
  basicstyle=\ttfamily\footnotesize,
  breaklines=true,
  xleftmargin=2em,
  numbers=left,
  keywordstyle=\color{blue},
  stringstyle=\color{mGreen},
  commentstyle=\color{mGray}
}

\lstdefinestyle{markdown}{
  basicstyle=\ttfamily\scriptsize,
  breaklines=true,
  xleftmargin=1.5em,
  numbers=none
}

\usepackage[most]{tcolorbox}
\newtcolorbox{findingbox}{
  colback=gray!10,
  colframe=black!70,
  arc=3pt,
  boxrule=0.5pt,
  left=6pt, right=6pt, top=4pt, bottom=4pt,
}

\definecolor{PromptBlue}{RGB}{58,78,105}
\definecolor{PromptBackground}{RGB}{244,247,250}
\newtcolorbox{promptbox}[1]{
  enhanced,
  breakable,
  colback=PromptBackground,
  colframe=PromptBlue,
  colbacktitle=PromptBlue,
  coltitle=white,
  title={#1},
  fonttitle=\bfseries,
  arc=4pt,
  boxrule=0.6pt,
  left=6pt, right=6pt, top=5pt, bottom=5pt,
  before skip=4pt, after skip=4pt,
}

\newcommand*\circled[1]{\tikz[baseline=(char.base)]{
            \node[shape=circle,fill,inner sep=1pt] (char) {\textcolor{white}{#1}};}}

\begin{document}

\date{}

\title{Beyond the Editing Canvas: Evidence Divergence in
OOXML-to-LLM Ingestion}
\author{
{\rm Side Liu}$^{1}$ \quad
{\rm Jiangpeng Liu}$^{1}$ \quad
{\rm Jinwen Xin}$^{2}$ \quad
{\rm Guojun Peng}$^{2}$ \quad
{\rm Jiang Ming}$^{1}$\\
$^{1}$Tulane University \qquad $^{2}$Wuhan University
}
\hypersetup{
  pdftitle={Beyond the Editing Canvas: Evidence Divergence in OOXML-to-LLM Ingestion},
  pdfauthor={Side Liu, Jiangpeng Liu, Jinwen Xin, Guojun Peng, and Jiang Ming}
}

\maketitle
\input{pages/0.abstract}
\input{pages/1.introduction}
\input{pages/2.background}
\input{pages/4.threatmodel}

\input{pages/5.methodology}

\input{pages/6.evaluation}
\input{pages/7.defense}

\input{pages/7.discussion}

\input{pages/8.conclusion}

{
\bibliographystyle{unsrt}
\bibliography{refs/office,refs/online,refs/fuzzing,refs/pdfs}
}
\input{pages/9.ethics}

\end{document}

%% file: pages/0.abstract.tex
\begin{abstract}

LLM pipelines increasingly ingest Office Open XML (OOXML) documents (Word, Excel, and PowerPoint files) as first-class evidence in financial, compliance, and retrieval-augmented workflows, implicitly assuming semantic integrity:
that the evidence consumed by the model matches the content shown in the
Microsoft Office suite editing canvas. We show that this assumption can fail in
OOXML-to-LLM pipelines.
The same specification-valid OOXML file can yield one evidentiary view in
Microsoft Office and another when extracted for an LLM. Each view is treated as
authoritative by its consumer, a condition we call \emph{plural ground truth}.
The ingestion contract rarely states which view and semantic roles become model
evidence or preserves how that evidence was derived. We call the
specification-grounded OOXML constructions that induce such divergence
\emph{evidence forks}.

We systematically traverse and mine the OOXML specification and confirm 21
evidence forks across Excel, Word, and PowerPoint, spanning six dimensions of
view construction. All 13 tools in our extraction panel emit evidence from at
least one fork. We test four native-ingestion LLM APIs and seven web chatbots.
Each test document carries a \emph{trap}: a task-relevant fact exposed by
extraction but not shown in Office. Across this 21-mechanism evaluation, the four
APIs return the trap in 48--76\% of trials. For 20 of 21 mechanisms, at least one of the eleven
interfaces returns the trap.
Our measurements further show that exposure is shaped upstream of the model by
the ingestion path and extractor configuration. A source-level survey of sixteen
popular open-source LLM projects further shows that default OOXML ingestion paths
concentrate on affected extractor families.

\end{abstract}

%% file: pages/1.introduction.tex
\section{Introduction}
\label{introduction}

Word, Excel, and PowerPoint files (\texttt{.docx}, \texttt{.xlsx}, and
\texttt{.pptx}) are ubiquitous~\cite{statista365} and increasingly serve as
inputs to large language model (LLM) applications~\cite{naveed2025comprehensive}.
Retrieval-augmented generation (RAG) systems load them into searchable knowledge
bases~\cite{langchain,llamaindex}, while financial agents use spreadsheets and
reports to support analysis and decisions~\cite{spreadsheetllm,dong2025finagents}.
These workflows rely on an \emph{ingestion} step that users do not see: a
backend parser, converter, or loader turns an Office Open XML (OOXML) file into
text for the model. They therefore assume \emph{semantic integrity}: the model
receives the same content that the user reviewed in Microsoft Office's default
editing canvas. When this assumption fails, the opaque ingestion layer becomes
a document supply-chain vulnerability.

Prior work on PDF documents has studied render/extract divergence, showing that
content masking can make online services act on evidence that readers do not
see~\cite{markwood2017pdfmirage}. Recent work on PDF-to-LLM pipelines examines
how content masking, phantom tokens, glyph remapping, reading order, and layout
shape the evidence delivered to LLMs
\cite{phantompdf2026,jin2025trapdoc,xiong2025invisible}. OOXML exposes a different,
more structural surface. It encodes values together with formulas, cached results,
formatting rules, application state, metadata, and compatibility
markup~\cite{ecma376,iso29500}.
For example, an \texttt{.xlsx} cell can carry a formula, its cached result, and a
number format that changes what Excel displays. OOXML defines the roles of these
representations, but an OOXML-to-LLM ingestion contract rarely specifies whether model
evidence should reflect the recalculated, cached, stored, or displayed value. What a
user sees in Microsoft Excel can therefore differ from what an LLM receives after
extraction.

OOXML document security research has exposed dangerous specification
features~\cite{muller2020office}, signature and rendering
gaps~\cite{rohlmann2023signatures}, ZIP-parser
differentials~\cite{you2025zip}, and hidden content in package structure and
revision data~\cite{castiglione2011stego}. Recent LLM-facing studies have
also tested hidden and off-page Office content against loaders, grading systems,
and web-assisted review workflows
\cite{castagnaro2025hidden,sili2025universal,tomazini2026hidden}. Together,
these studies establish that selected Office carriers can survive ingestion and
affect downstream tasks. Their methodological starting point is a predefined
set of document manipulations, whose effects they then measure in a target
workflow.

In this paper, we start from the OOXML specification and systematically study
how \emph{plural ground truth} can arise from a single specification-valid OOXML
file: Microsoft Office's default editing canvas and an ingestion pipeline expose
incompatible task-relevant evidence. Each consumer treats its own view as
authoritative; this operational plurality does not imply normative equivalence
under OOXML. This violates the semantic-integrity assumption that a
model reasons over the evidence shown in the review view. We call a
specification-grounded OOXML construction that induces this evidentiary
divergence an \emph{evidence fork}. The fork arises during view construction:
the stored package is unchanged, but Office and the ingestion path promote
different content to task evidence.

In all, we confirm 21 evidence forks across
Word, Excel, and PowerPoint, spanning representation, state, compatibility,
visibility, scope, and linearization. Across 13
OOXML extraction tools, mechanism reach is sharply bimodal: seven mechanisms reach
11--33\% of applicable tools and fourteen reach 67--100\%, with none in between. The
split follows where the trap resides in the OOXML package: ordinary content is broadly
exposed, whereas attributes, relationships, and alternate branches are tool-specific.
We build ten realistic, specification-valid Office documents per mechanism from
TAT-QA financial-report excerpts~\cite{zhu2021tatqa}, 210 in total, and test them
against four native-ingestion LLM APIs and seven mainstream web chatbots. Each
instance carries a \emph{trap}: a task-relevant fact exposed by extraction but not
shown in Office. Across 8{,}400 API trials, models return the trap in 48--76\% of
runs, depending on the API. At least one interface returns the trap for 20 of 21
mechanisms. Exposure varies with
the ingestion path even across interfaces that a provider identifies as using
the same model. A service's exposure pattern can also match it
to candidate ingestion configurations that behave identically on our probes. A
source-level survey of sixteen popular open-source LLM projects shows that their
default OOXML ingestion paths concentrate on affected extractor families.

To our knowledge, this is the first systematic, specification-grounded
measurement of divergence between the default Office view and the task evidence
delivered by OOXML-to-LLM pipelines. Our study covers Word, Excel, and
PowerPoint across heterogeneous extractors and LLM interfaces.
We make three contributions:
\begin{itemize}[leftmargin=1.4em,itemsep=2pt,topsep=2pt]
    \item We introduce a specification-guided method for discovering and
    confirming evidence forks, yielding a unified cross-format catalog and a
    six-dimensional taxonomy of OOXML view constructions.
    \item We conduct a large-scale measurement of how evidence forks propagate
    through extraction tools and LLM interfaces. We find broad
    exposure across the tested systems and show that it is shaped by ingestion
    path and extractor configuration, not by model choice alone.
    \item We assess ecosystem reach and detectability across open-source LLM
    projects and ordinary documents, revealing shared extractor dependencies
    and the limits of presence-only screening for common OOXML channels.
\end{itemize}

%% file: pages/2.background.tex
\section{Background}
\label{sec:background}

This section provides the document-format context for our study. We first
describe how OOXML packages organize document content and supporting state,
then use a formula-cache example to illustrate how the same file can yield
different evidence across consumers. We next introduce the Office applications
and extraction tools that construct these views. Finally, we position our work
relative to prior research on Office-document security and document attacks on
LLMs.

\subsection{OOXML Documents}

\noindent\textbf{OOXML Structure }
Office Open XML (OOXML) is the file format of the common Office documents,
mainly \texttt{.docx}, \texttt{.xlsx}, and \texttt{.pptx}. Each is an Open Packaging
Conventions (OPC) container: a ZIP archive of typed \emph{parts} (XML documents
and media) tied together by \emph{relationships}~\cite{ecma376,iso29500}.
Listing~\ref{lst:docx} shows the internal structure of a real Word \texttt{.docx}:
a main story, supporting parts, and the relationships that connect them. The
visible text is the main story in \texttt{word/document.xml}, while its styles,
numbering, fonts, and recalculation and compatibility settings live in sibling
parts; package-level parts such as \texttt{[Content\_Types].xml} and
\texttt{\_rels/.rels} declare each part's type and record the relationships. Beyond
these, a document may carry further content such as comments, footnotes, headers
and footers, and drawing objects, each in a part that a given consumer may or may
not traverse. User-facing applications resolve these parts through rendering,
recalculation, style application, and data binding. Text extraction tools, by contrast, linearize the
package through partial parsers that prioritize speed, portability, or text
coverage over faithful application-level resolution.

\vspace*{2pt}
\begin{lstlisting}[numbers=none,basicstyle=\ttfamily\small,breaklines=false,frame=single,xleftmargin=4pt,aboveskip=4pt,belowskip=2pt,captionpos=b,label={lst:docx},caption={Inside a real \texttt{.docx}: the OPC package of typed parts, with the \texttt{word/document.xml} body story.}]
example.docx
 |-- [Content_Types].xml   %content-type map
 |-- _rels/.rels           %package relationships
 |-- docProps/             %core.xml, app.xml, ...
 |-- customXml/            %item1.xml  (data-bound)
 `-- word/
     |-- document.xml      %main story (body text)
     |-- styles.xml        %styles + number formats
     |-- settings.xml      %recalc + compatibility
     |-- theme/, numbering.xml, fontTable.xml ...
     `-- _rels/document.xml.rels

word/document.xml:
  <w:body>
    <w:p><w:r><w:t>Topic 606 ...</w:t></w:r></w:p>
    ...
  </w:body>
\end{lstlisting}

\begin{figure}[]
\centering
\includegraphics[width=\columnwidth]{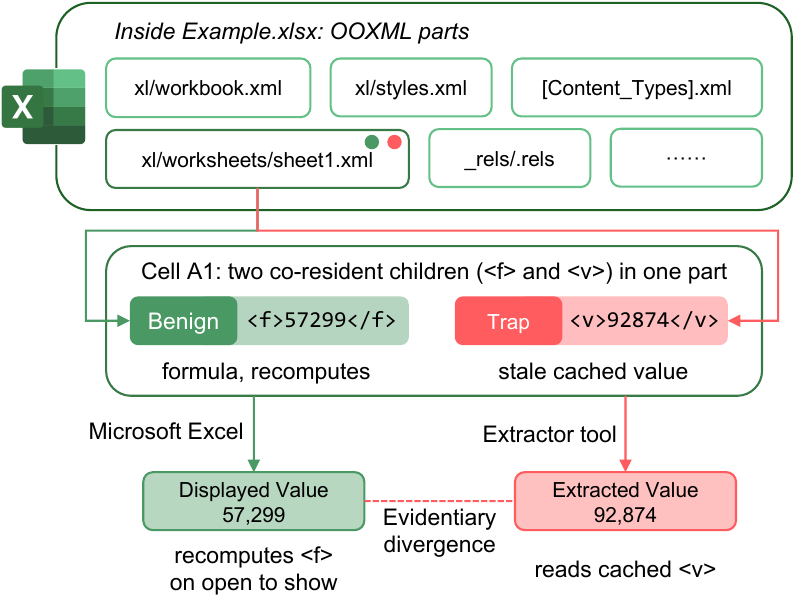}
\caption{A formula/cache evidence fork: Microsoft Excel shows the benign value $57{,}299$
while an extractor feeds the trap value $92{,}874$ to the model.}
\label{fig:motivation}
\end{figure}

\vspace*{2pt}
\noindent\textbf{Motivation Example }
The same object in an OOXML file can carry multiple representations with
different semantic roles, and an Office editor and an extractor can expose
different ones. Figure~\ref{fig:motivation} shows the simplest case. A worksheet cell
stores a formula \texttt{<f>} next to its cached
result \texttt{<v>}: Microsoft Excel recomputes the formula on open and shows the
benign value 57{,}299, while an extractor without a formula engine reads the
attacker-planted cache \texttt{<v>} of 92{,}874. Neither reading
is hidden or malformed, yet the two consumers disagree.

OOXML permits the formula and its cached result to coexist and assigns them
distinct roles. Excel recalculates and renders the formula result, whereas a
formula-less extractor may expose the cache directly. The ingestion contract does
not state which of these representations becomes model evidence or record that the
emitted value came from a cache. The specification-valid file opens cleanly and
shows the benign 57{,}299 in Excel's default editing canvas, with no macro to run
and no part to repair. An attacker who controls the cell can make the model ingest
the trap 92{,}874 from the extractor while the analyst reviews the benign value.
This formula/cache construction is a representation-level evidence fork; the
difference between the two outputs is the resulting evidentiary divergence. The
remaining classes arise from other decisions involved in constructing a document
view.

\subsection{OOXML Consumers}

\noindent\textbf{Office Applications }
A user opens and reviews an Office document in an office application that renders it,
chiefly the Microsoft Office suite~\cite{msoffice} (Word, Excel, PowerPoint) or
LibreOffice~\cite{libreoffice}. Such an application recalculates formulas, applies
number formats and styles, refreshes data bindings, resolves compatibility branches,
and hides filtered or hidden content. We therefore choose the Microsoft Office suite's
default editing canvas as the human-facing oracle in our threat model, reflecting a
widely deployed OOXML implementation and the review workflow we study.

\vspace*{2pt}
\noindent\textbf{Extraction Tools }
When the same document reaches an LLM, its text is produced not by such an application
but by a separate extraction tool that runs first. These tools are libraries:
format-specific parsers such as \texttt{python-docx}~\cite{pythondocx},
\texttt{openpyxl}~\cite{openpyxl}, and \texttt{python-pptx}~\cite{pythonpptx};
converters such as \texttt{docx2txt}~\cite{docx2txt}, \texttt{xlsx2csv}~\cite{xlsx2csv},
\texttt{pandoc}~\cite{pandoc}, Apache Tika~\cite{apachetika}, and
LibreOffice~\cite{libreoffice}; and meta-extractors such as markitdown~\cite{markitdown},
Unstructured~\cite{unstructured}, and Docling~\cite{docling} that wrap the others. Each
is a \emph{partial} parser that linearizes the package for speed, portability, or text
coverage rather than reproducing an application's resolution. Across mainstream
open-source RAG and agent projects (LangChain, LlamaIndex,
Haystack, AutoGen, Unstructured, Docling, markitdown, RAGFlow, Dify, Onyx, and txtai,
among others), default OOXML ingestion paths concentrate on the same small set of leaf
extractors, making loader choice an indicator of potential exposure.

\subsection{Related Work}
\label{sec:relatedwork}

\noindent\textbf{Office-Document Security }
M\"uller et al.~\cite{muller2020office} systematically read the OOXML and ODF
specifications to catalog dangerous-by-design features and application-level
content masking. Other work studies signature and rendering
gaps~\cite{rohlmann2023signatures}, container and parser
differentials~\cite{you2025zip}, hidden data in package structure and revision
state~\cite{castiglione2011stego}, and executable macro
semantics~\cite{ruaro2022symbexcel}. These studies target Office applications,
hosts, signatures, or forensic recovery. We instead study the evidence supplied
to LLMs; our fixtures are specification-valid and open without repair, and
malformed containers and macros are outside our threat model.

\vspace*{2pt}
\noindent\textbf{Document Attacks on LLMs }
Recent PDF-to-LLM attacks use phantom tokens, glyph remapping, reading-order
manipulation, and layout to make models act on evidence that differs from a
page's rendered content
\cite{phantompdf2026,jin2025trapdoc,xiong2025invisible}. The closest
LLM-facing Office document studies examine selected hidden and off-page carriers in
Word documents. Castagnaro et al.~\cite{castagnaro2025hidden} evaluate six
techniques applicable to DOCX against loaders and RAG systems.
S{\^\i}li et al.~\cite{sili2025universal} test four predefined DOCX injection
techniques in LLM-based grading. Tomazini et al.~\cite{tomazini2026hidden} use
a single white-font Word carrier to alter web-LLM review recommendations.
These studies establish carrier-level and task-level effects in DOCX, but do
not examine Excel or PowerPoint. We instead derive 21 OOXML view constructions
from the specification across Word, Excel, and PowerPoint and trace them
through heterogeneous extraction tools and LLM interfaces.
Prompt-in-content and indirect-prompt-injection studies place attacker
instructions in documents or other structured inputs to redirect model
behavior~\cite{lian2025promptincontent,greshake2023notwhat,feng2025struphantom}.
Other work uses LLMs to detect malicious OOXML files~\cite{hess2025detection},
develops detectors for hidden prompts~\cite{murray2025phantomlint}, and measures real-world
prompt injection in resume-screening workflows~\cite{zhang2026resume}.

%% file: pages/4.threatmodel.tex
\subsection{Threat Model}
\label{sec:threatmodel}

\begin{figure*}[t]
\centering
\includegraphics[width=\textwidth]{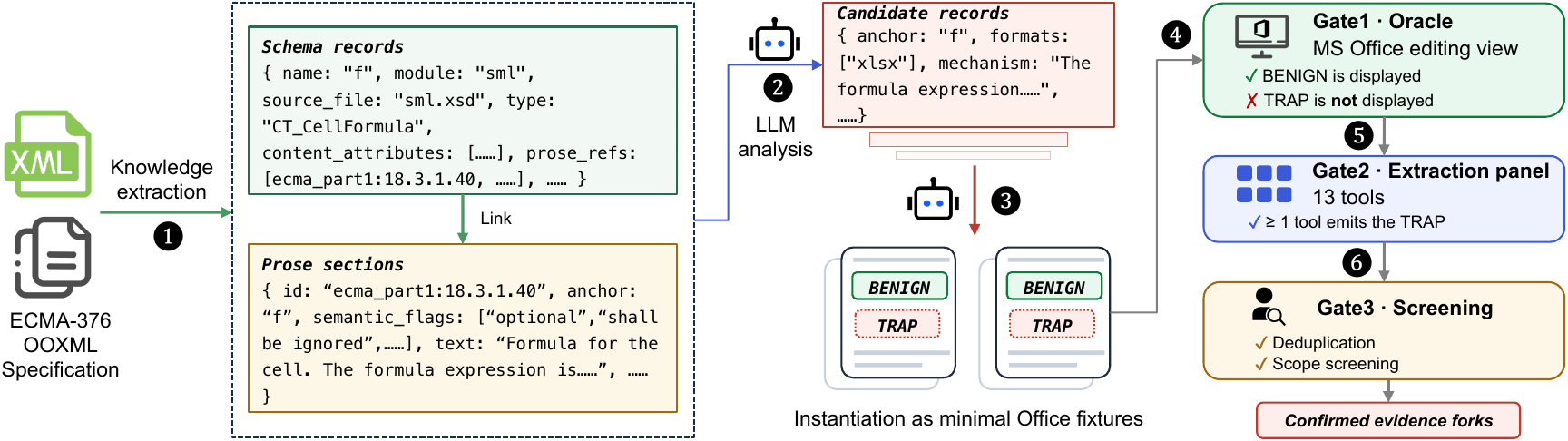}
\caption{Overview of our workflow for discovering evidence forks from the OOXML
specification.}
\label{fig:pipeline}
\end{figure*}

We consider an OOXML-to-LLM workflow in which a user reviews a document in
Microsoft Office's default editing canvas, while an ingestion pipeline
independently extracts evidence from the same file for factual question answering
or summarization. The attacker supplies the document and controls its OOXML
content and package structure, but not the extractor, model, or prompt. The file
must be specification-valid, open in Word, Excel, or PowerPoint without repair,
display the benign content, and not display the trap in the default editing
canvas. We exclude malformed containers, macros, and executable payloads.
The attack succeeds when the LLM returns attacker-chosen, task-relevant content
that is absent from the Office review view, either by substituting a queried fact
or adding a fabricated disclosure. We model ordinary review in the default
editing canvas, not forensic inspection through auxiliary views or package
internals.

%% file: pages/5.methodology.tex
\section{Workflow}
\label{sec:methodology}

We identify evidence forks using the workflow in Figure~\ref{fig:pipeline}. First,
we turn the OOXML specification into a queryable knowledge base
(step~\circled{1}), then traverse it with an LLM to produce candidate records
(\circled{2}). We attempt to instantiate each record as a minimal Office fixture
(\circled{3}) and test it with the Office oracle (\circled{4}) and extraction
panel (\circled{5}). Finally, we deduplicate passing records into mechanisms and
apply scope screening (\circled{6}), yielding the confirmed catalog.

\subsection{Knowledge Base and Mining}
\label{sec:mining}

The OOXML specifications combine machine-readable schemas with extensive prose.
We organize both into a queryable knowledge base (\circled{1}) and mine it with
an LLM agent to produce candidate records (\circled{2}).

\vspace*{2pt}
\noindent\textbf{Knowledge Base Construction }
The OOXML specification comes in two forms. The \emph{prose} reference---ECMA-376~\cite{ecma376,iso29500} and
the Microsoft extension documents~\cite{msdocx,msxlsx,mspptx}, which describe each element in
numbered clauses and states the rules a parser must obey.
The \emph{XSD} (XML Schema Definition) schemas~\cite{openxmlsdk} give
the same elements a machine-readable grammar: what each element may contain, its
attributes, and their multiplicities.
We build one index from each.
Parsing the schemas yields 15{,}884 schema records, one per element or
type, each carrying its content model, \texttt{xs:choice} siblings, occurrence,
defaults, and references to its associated prose clauses. Cutting the prose at
its clause headers yields 5{,}206 prose sections, each anchored to an
element name and assigned an identifier that preserves its specification source
and section number. Both are emitted as JSON. Schema records store these
source-qualified prose-section identifiers in \texttt{prose\_refs}, keeping
same-named clauses distinct.

\noindent\textbf{Mining }
We traverse the knowledge base with an LLM (Claude Opus~4.8). The base is too
large for one context, and per-element attention degrades in very long inputs, so we
partition it into element-anchored chunks of at most 500K characters that never split a
section---23 chunks across the eight blocks: the ECMA-376 and the Microsoft extensions. Each chunk lists its elements as
\texttt{[name] title (\S section)} followed by the element's full clause---its attribute
table and any precedence, selection, or visibility rule. For every chunk we issue one
agent call that goes element by element and, whenever an OOXML construct could cause
Office rendering and text extraction to expose different task-relevant content, emits
a candidate record with the following fields:

\begin{quote}\small
\texttt{(element, \S section, format, quote, render, extract, strength)}
\end{quote}

\noindent Each field feeds a later stage. \texttt{element} and \S\texttt{section} anchor
the record to a construct, support anchor-level consolidation, and locate it within the
traversed specification blocks;
\texttt{format} selects the document type to instantiate; \texttt{quote} is the verbatim
specification sentence that defines the construct, its semantic role, or its processing
rule, providing an audit trail; \texttt{render} and \texttt{extract} describe the
hypothesized Office and extractor outputs, which become the \textsc{benign} and
\textsc{trap} tokens during empirical instantiation (\circled{3}); \texttt{strength}
records how directly the specification supports the record. Each call writes its
tuples to a result file. A deterministic parser merges the 23 result files and
consolidates records sharing the same specification anchor on \texttt{(element, \S section)}, yielding 639
specification-anchored candidate records. At this stage, each record is an observation
tied to one specification location, not yet a distinct mechanism.

\subsection{Instantiation and Confirmation}
\label{sec:fixtures}\label{sec:oracle}\label{sec:panel}

The candidate records are not yet confirmed evidence forks: the
specification establishes the construct and its semantics, but not how an Office
application or extractor operationalizes it. Confirmation therefore uses two
behavioral tests: the default Office editing canvas must show
\textsc{benign} rather than \textsc{trap}, and at least one extractor must emit
\textsc{trap}.
We attempt to instantiate all 639 candidate records as minimal,
specification-valid Office fixtures (\circled{3}) and evaluate them with the Office
oracle (\circled{4}) and extraction panel (\circled{5}). Passing records then undergo
mechanism-level deduplication and scope screening (\circled{6}) to form the confirmed
catalog.

\vspace*{2pt}
\noindent\textbf{Instantiation }
To keep instantiation auditable, we ask an LLM to generate a short builder program
rather than package bytes directly. Using a builder makes every package edit explicit
and reproducible. It also lets us distinguish construction failures from mechanism
failures before applying the two behavioral gates. We
prompt the model with its supporting candidate record from the mining stage: the element and
section, the format, the verbatim specification quote, the value Office renders, and the
value an extractor reads. From these it writes a Python builder that produces the
smallest Office file exercising that one construct. The builder uses format-specific
Python libraries (\texttt{python-docx}, \texttt{openpyxl}, \texttt{python-pptx}) where
the construct is exposed, and edits the underlying part XML in place where it is not (a
\texttt{webHidden} run, an off-page \texttt{framePr}, an \texttt{mc:AlternateContent}
branch). It places a unique \texttt{BENIGN} token in the Office-displayed view and a
unique \texttt{TRAP} token in the alternative representation or channel. These
distinct tokens let us test their presence or absence precisely in Office screenshots
and extractor outputs.
Within a three-attempt construction budget, 542 yield runnable fixtures;
the remaining 97 either lack a suitable carrier for the controlled
tokens or cannot be instantiated within that budget.
Each successful run yields a candidate fixture for the two behavioral gates
and Gate~3 screening.

\vspace*{2pt}
\noindent\textbf{Confirmation }
The confirmation stage consists of two behavioral gates followed by a deduplication and
scope-screening.

\emph{Gate~1, the oracle}, fixes what the default editing canvas shows: it must display
the \texttt{BENIGN} token and not the \texttt{TRAP}. We automate this
with a script that opens the file in Microsoft Office and captures its editing view. A
multimodal LLM then reports whether the \texttt{BENIGN} and \texttt{TRAP} tokens are
visible in the image. The judgment reduces to checking whether two distinct,
attacker-chosen strings are present in a rendered screenshot rather than performing
general visual recognition or reasoning. A candidate fixture passes Gate~1 only when
\texttt{BENIGN} is visible and \texttt{TRAP} is not.
After Gate~1, 412 candidate fixtures remain and proceed to Gate~2.

\emph{Gate~2, the extraction panel}, requires that at least one extraction tool emit the
\texttt{TRAP}. Table~\ref{tab:panel} lists the 13 tools, chosen to span what real
ingestion stacks rely on: format-specific libraries, command-line converters, and the RAG
loaders that front many LLM pipelines. One tool suffices: a single extractor that
surfaces the trap makes that view available to the downstream model. We look up whether each trap token
appears in a tool's output after normalizing whitespace, rather than matching the whole
string, so that harmless formatting differences in the \texttt{BENIGN} reading are not
mistaken for a divergence. A candidate fixture passes Gate~2 when some tool emits the
\texttt{TRAP}.
After Gate~2, 163 candidate fixtures remain.

\begin{table}[]
\centering
\footnotesize
\setlength{\tabcolsep}{5pt}
\caption{The extraction tools panel.}
\label{tab:panel}
\begin{tabular}{llccc}
\toprule
Tool  & Version & xlsx & docx & pptx \\
\midrule
openpyxl ~\cite{openpyxl}  & 3.1.5      & \CheckmarkBold & & \\
python-calamine~\cite{calamine}     & 0.7.0      & \CheckmarkBold & & \\
xlsx2csv~\cite{xlsx2csv}            & 0.8.6      & \CheckmarkBold & & \\
python-docx~\cite{pythondocx}       & 1.2.0      & & \CheckmarkBold & \\
docx2txt~\cite{docx2txt}            & 0.9        & & \CheckmarkBold & \\
python-pptx~\cite{pythonpptx}       & 1.0.2      & & & \CheckmarkBold \\
pandoc~\cite{pandoc}                & 3.10       & & \CheckmarkBold & \\
textutil                            & macOS~26.1 & & \CheckmarkBold & \\
Apache Tika~\cite{apachetika}       & 2.9.2      & \CheckmarkBold & \CheckmarkBold & \CheckmarkBold \\
LibreOffice~\cite{libreoffice}      & 26.2.4.2   & \CheckmarkBold & \CheckmarkBold & \CheckmarkBold \\
markitdown~\cite{markitdown}        & 0.1.6      & \CheckmarkBold & \CheckmarkBold & \CheckmarkBold \\
Unstructured~\cite{unstructured}    & 0.23.1     & \CheckmarkBold & \CheckmarkBold & \CheckmarkBold \\
Docling~\cite{docling}              & 2.108.0    & \CheckmarkBold & \CheckmarkBold & \CheckmarkBold \\
\bottomrule
\end{tabular}
\end{table}

\emph{Gate~3} applies deduplication and scope screening to the candidates that pass
Gates~1 and~2. We remove 104 duplicate records that exercise the same package-level
construction.
We retain candidates whose divergence follows from an OOXML-defined state, content role,
or processing rule. We exclude purely visual camouflage, such as white-on-white text,
object overlap, and extremely small text, because Office still renders the underlying
content. We also exclude PowerPoint speaker notes, which Office presents in a
dedicated notes area rather than as part of the slide, and reversible encodings
that leave the trap visible. These scope criteria exclude 38 candidates, leaving
21 distinct mechanisms in the confirmed catalog.

\vspace*{2pt}
\noindent\textbf{The Confirmed Evidence Forks }
The confirmation workflow yields 21 evidence forks. Table~\ref{tab:mechanism_map}
maps the identifiers used throughout the evaluation to short names, document formats,
and view-construction classes. The detailed catalog in
Appendix~A identifies each OOXML construct in a specification-valid
fixture and its empirically observed Office-versus-extractor divergence. We organize
the mechanisms by the OOXML processing point at which the views diverge;
Table~\ref{tab:view_dimensions} summarizes these six dimensions.

\begin{table}[t]
\centering
\footnotesize
\setlength{\tabcolsep}{5pt}
\caption{Compact map of the mechanism identifiers used throughout the evaluation.
Complete construct-level descriptions appear in Appendix~A.}
\label{tab:mechanism_map}
\begin{tabular}{p{0.22\columnwidth} l l p{0.45\columnwidth}}
\toprule
\textbf{Class} & \textbf{ID} & \textbf{Fmt} & \textbf{Short name} \\
\midrule
\multirow{5}{=}{Representation}
 & M1 & xlsx & Fixed-literal \texttt{numFmt} \\
 & M2 & xlsx & Blank \texttt{numFmt} \\
 & M3 & xlsx & Conditional-format override \\
 & M4 & xlsx & Formula / stale cache \\
 & M5 & docx & Binding / cached run \\
\midrule
\multirow{6}{=}{State}
 & M6  & xlsx & Hidden sheet \\
 & M7  & xlsx & Very-hidden sheet \\
 & M8  & xlsx & Filter-hidden rows \\
 & M9  & xlsx & Collapsed outline rows \\
 & M10 & xlsx & Hidden column \\
 & M11 & xlsx & Normal-view header \\
\midrule
\multirow{3}{=}{Visibility}
 & M12 & docx & Vanished run \\
 & M13 & pptx & Hidden shape \\
 & M14 & docx & Off-page text box \\
\midrule
\multirow{2}{=}{Compatibility}
 & M15 & docx & Unsatisfied MC Choice \\
 & M16 & docx & Unselected MC Fallback \\
\midrule
\multirow{3}{=}{Scope}
 & M17 & docx & Drawing alt text \\
 & M18 & docx & Table description \\
 & M19 & docx & Hyperlink display / target \\
\midrule
\multirow{2}{=}{Linearization}
 & M20 & xlsx & Covered merged-cell value \\
 & M21 & xlsx & Phonetic guide \\
\bottomrule
\end{tabular}
\end{table}

\emph{Representation} (five) covers objects whose displayed or computed value is not the
value stored: a custom number format rewrites the cell, an always-true conditional format
overrides it, a formula recomputes over a stale cache, or a content control refreshes a
data binding, while an extractor returns the raw stored value. \emph{State} (six) hides
content behind a container or view state that is present but not shown: a hidden or
very-hidden sheet, an autofilter- or outline-collapsed row, a hidden column, a worksheet
header absent from Normal view. \emph{Visibility} (three) suppresses an object that would
otherwise render, through a vanished run, a hidden shape, or a text box pushed off the
page. \emph{Compatibility} (two) covers Markup-Compatibility selection: Office retains
the applicable branch, while an MC-blind extractor may emit content from an unselected
branch. \emph{Scope} (three) emits a
non-body field as content: drawing alt text, a table description, a hyperlink target.
\emph{Linearization} (two) flattens structure, so that a value hidden under a merged cell,
or a phonetic guide beside its base run, reaches the extractor out of place. State and
representation are the largest classes, at six and five mechanisms; the rest hold two or
three. By document type, twelve of the 21 occur in Excel, eight in Word, and one in
PowerPoint.

Representation mechanisms are particularly important because several do not rely
on hidden or off-canvas content. A number-format or conditional-format override
(M1, M3), a formula paired with a stale cache (M4), or a data binding (M5)
places co-resident values or roles on one object. Office and an extractor can
therefore expose different task evidence without suppressing or moving text. A
search limited to hidden content would miss this structured part of the attack
surface.

\begin{table}[]
\centering
\footnotesize
\setlength{\tabcolsep}{3pt}
\caption{Six dimensions of OOXML view construction.}
\label{tab:view_dimensions}
\begin{tabular}{p{0.21\columnwidth}p{0.73\columnwidth}}
\toprule
\textbf{Class} &
\textbf{View-construction step} \\
\midrule
Representation &
Which encoded representation supplies an object's value \\
\addlinespace[1pt]

State &
Whether encoded content is active in the current application view \\
\addlinespace[1pt]

Visibility &
Whether a realized object appears in the rendered view \\
\addlinespace[1pt]

Compatibility &
How Markup Compatibility branch selection is applied \\
\addlinespace[1pt]

Scope &
Which non-body fields count as document evidence \\
\addlinespace[1pt]

Linearization &
How structured content is flattened into text \\
\bottomrule
\end{tabular}
\end{table}

\subsection{A Realistic Test Set}
\label{sec:testset}
After confirming each fork on a minimal fixture, we embed it in realistic Office
documents for evaluation.
Content comes from TAT-QA~\cite{zhu2021tatqa}, a JSON corpus of hybrid table-and-text
excerpts from financial reports. For each sample, we render one excerpt as a new Excel,
Word, or PowerPoint file in the mechanism's target format and instantiate one evidence
fork. This yields ten samples per mechanism, 210 in all. In each, the value in the
Office-displayed view is the filing's true value, while the extractor-emitted value or
content is the attacker's trap. Injections are minimal, so each file remains
specification-valid and opens normally. Before its first web-chat upload, an author
opens each instance in the default Microsoft Office editing view used by Gate~1
and manually verifies that the source content renders normally and that the
planted content is not shown. All 210 instances pass this check. We separately validate
all files with \texttt{OpenXmlValidator} from Microsoft Open XML SDK~3.5.1~\cite{openxmlsdk},
targeting Microsoft~365. It reports no package, schema, semantic,
or Markup Compatibility errors for any file.

Trap placement follows two task families. In the
\emph{question-answering} family, the attacker uses the fork to substitute a single
queried fact: a cell whose
displayed value is the true figure but whose stored value reads a different number, or a
counterparty whose visible name differs from the one an extractor emits. In the
\emph{summarization} family, the attacker uses the fork to plant a fabricated
disclosure that is absent from
the page yet present to an extractor, so that a model asked to summarize the filing
reports a liability, provision, or related-party dealing the document never states.
Three of these samples, one XLSX, one DOCX, and one PPTX, are walked through in
Appendix~D. Each sample carries a manifest with its task
question, benign answer, and trap answer.

%% file: pages/6.evaluation.tex
\section{Measurement}
\label{sec:evaluation}

We evaluate evidence-fork propagation, ecosystem reach, and detectability
through five questions:
\begin{enumerate}[leftmargin=3em,topsep=2pt,itemsep=0pt,parsep=3pt,
                  label=\textbf{RQ\arabic*.}]
    \item Which extractors expose the trap-bearing view?~(\S\ref{sec:rq1})
    \item Do end-to-end LLM services propagate that view into model
    outputs?~(\S\ref{sec:rq2})
    \item Model or ingestion path: which sets exposure?~(\S\ref{sec:rq3})
    \item How widely do affected ingestion paths appear in open-source agent and
    RAG projects?~(\S\ref{sec:rq4})
    \item How common are these structures in ordinary files, and what do their
    benign base rates imply for presence-only screening?~(\S\ref{sec:rq5})
\end{enumerate}
\input{figs/tab_reach}
\subsection{Experimental Setup}
\label{sec:setup}

\noindent\textbf{Measurement Scope }
We evaluate the 21 evidence forks across extraction tools and LLM services on the 210
document instances of \S\ref{sec:testset}. The extraction
panel comprises 13 tools (Table~\ref{tab:panel}). We further test four
native-ingestion APIs, meaning they accept and parse an uploaded Office document
directly: GPT-5.5, Kimi K2.6, Qwen-doc-turbo, and GLM 5.2. Some mainstream APIs
such as Anthropic's Claude and Google's Gemini do not accept Office documents for
native parsing, so for
those models we use their web chat interface: we test seven web chatbots (Claude
Sonnet~5, ChatGPT GPT-5.5, DeepSeek V4, GLM 5.2, Gemini 3.5, Kimi K2.6, and Qwen-3.5)
to assess each mechanism's reachability and propagation in real deployments.

\vspace*{2pt}
\noindent\textbf{Trial Protocol }
For the API tests, we run each mechanism's ten instances ten times, 2{,}100 runs per
API and 8{,}400 trials in total. Across the seven web chatbots, we test each of
the 210 documents once, yielding 1{,}470 web trials.

\vspace*{2pt}
\noindent\textbf{Response Labeling and Validation }
We use Claude Opus~4.8 and GPT-5.5 to independently label all 8{,}400 API
responses under a binary rubric: \textsc{trap} means that a response asserts the
planted content as a document fact; all other responses are \textsc{non-trap}.
Appendix~C gives the complete prompt and input fields. The two
judges agree on all 8{,}400 responses (100\% raw agreement). We further audit five
randomly sampled responses from each mechanism--API combination (420 total). The
two authors split this audit set, and every human label agrees with both automated
verdicts. The two authors manually label and verify every web response during
testing; completing the 1{,}470 trials took about two weeks.

\vspace*{2pt}
\noindent\textbf{Controls }
We use two controls for the LLM measurement. First, we construct a matched benign
twin for each of the 210 documents instantiating an evidence fork by removing the divergent representation
or channel while preserving the source content and question, and query each API
once. Second, as a visible-positive sanity check, we construct six non-divergent
documents that place the designated target in ordinary visible cell, body, or slide
text. Together, these controls span the three OOXML formats and both task families
used in the main experiment.

\subsection{RQ1: Extractor Reachability}
\label{sec:rq1}

RQ1 measures extractor reachability. Table~\ref{tab:reach} gives the full matrix; each tool
runs at one pinned version with one set of call arguments, listed in Table~9 in
Appendix~B. Across the confirmed set, \emph{width} (the number of applicable
tools that emit the trap) ranges from a single tool to every applicable one. Each
mechanism--tool outcome is identical across its ten base documents, indicating
construct-stable behavior within our test set. The spread itself is
unsurprising: read as a whole, though, the matrix reveals three structural properties
that no single row or column shows.

\begin{figure}[t]
\centering
\includegraphics[width=\columnwidth]{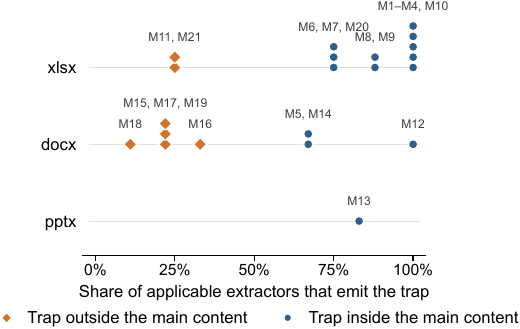}
\caption{Reach as a proportion of the tools that handle each format; each marker is one
mechanism.}
\label{fig:reachdist}
\end{figure}

\vspace*{2pt}
\noindent\textbf{The Format Decides Whether The Loader is a Security Choice }
Read as a proportion of the tools that handle each format, reach is sharply bimodal
(Figure~\ref{fig:reachdist}): seven mechanisms fall between 11\% and 33\%, the other
fourteen between 67\% and 100\%, and none lies in between. What separates the two bands is
where the trap is carried: every mechanism in the low band sits outside the main content,
in an XML attribute, a relationship target, an alternate-content branch, a worksheet
header, or a phonetic annotation, while every mechanism in the high band is carried as
ordinary content text that any walk of the document reaches. That is also what divides the
two formats. A spreadsheet keeps almost all of its mechanisms in the cell grid, which
retains visible and hidden cells in the same cell stream, so most extractors traverse
both: the three native spreadsheet tools read seven mechanisms in common and differ on at
most three, and five of the twelve spreadsheet mechanisms are read by every applicable
tool. Word instead scatters its mechanisms across attributes, branches, and nested stories,
and each library draws its own line on how much of the document tree to emit, so the four
native Word extractors share exactly one mechanism, the vanished run, and agree on nothing
else. The regularity holds inside a format as well as across formats, since the two
spreadsheet mechanisms that live outside the grid fall to 25\%, the same band as the Word
attribute channels. The consequence is asymmetric: loader choice has little effect
on high-reach XLSX mechanisms but substantially changes DOCX exposure.

\vspace*{2pt}
\noindent\textbf{Class and Reach Move Together }
Where a trap is carried is not an incidental property of a mechanism, it is what the classes
of \S\ref{sec:methodology} are defined by. \emph{Scope} means a non-body field is emitted as
content, so its trap necessarily sits outside the main content; \emph{compatibility} means a
Markup-Compatibility branch is selected, so its trap necessarily sits in an alternate
branch; \emph{representation} means the displayed value differs from the stored one, so its
trap is necessarily a stored value in the main content. The class fixes the carrier, and reach
follows the carrier closely. Table~\ref{tab:classreach} bears this out: ordering the six classes by
mean reach reproduces exactly their ordering by how much of each class sits inside the main
content, from representation at 0.93 with all five mechanisms inside, down to scope at 0.19
with none. The two orderings coincide across all six classes, so the classes group
mechanisms that behave alike at the extraction layer rather than merely sorting them for
presentation.

\begin{table}[]
\centering
\footnotesize
\setlength{\tabcolsep}{5pt}
\renewcommand{\arraystretch}{1.1}
\caption{Mean extractor reach by class, against the share of each class whose trap is
carried inside the main content.}
\label{tab:classreach}
\begin{tabular}{l c c c}
\toprule
\textbf{Class} & \textbf{Mech.} & \textbf{Mean reach} & \textbf{Inside main content} \\
\midrule
Representation  & 5 & 0.93 & 100\% \\
Visibility      & 3 & 0.83 & 100\% \\
State           & 6 & 0.75 & 83\% \\
Linearization   & 2 & 0.50 & 50\% \\
Compatibility   & 2 & 0.28 & 0\% \\
Scope           & 3 & 0.19 & 0\% \\
\bottomrule
\end{tabular}
\end{table}

\noindent\textbf{Broader Coverage Expands Exposure }
The one Word mechanism read by every applicable extractor is M12, the vanished
run. The other Word mechanisms are visible to some extractors and invisible to
others. \texttt{python-docx} reads only M12 because it does not surface bindings,
alternate branches, or accessibility attributes. General-purpose converters
designed for broader content coverage expose more mechanisms.
The two broadest panel members, \texttt{markitdown} and Apache Tika, expose 17
and 16 mechanisms. Their flattened outputs surface more OOXML channels without
preserving their Office-assigned roles. Text coverage and semantic integrity
therefore pull against each other, and a pipeline that upgrades its loader for
better coverage can silently widen its attack surface.

\begin{findingbox}
\textbf{Finding I.} Broader text traversal does not imply more faithful
view construction: broad converters expose more trap-bearing channels, while
narrow loaders can appear safer simply by omitting document semantics. Extractor
consensus therefore reflects shared traversal behavior, not agreement with the
Office view.
\end{findingbox}

\begin{figure}[]
\centering
\includegraphics[width=\columnwidth]{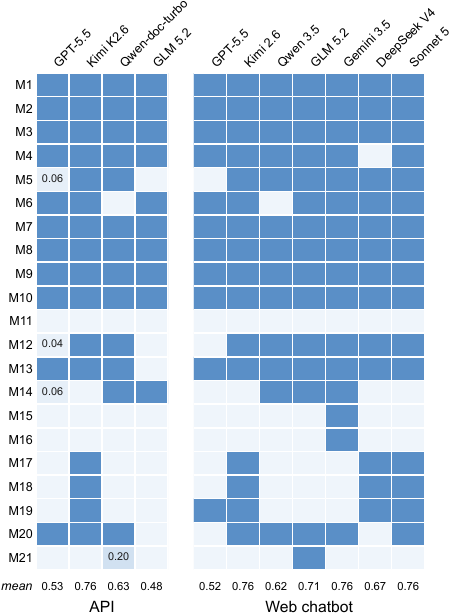}
\caption{Per-mechanism propagation into LLM output. Each cell is the fraction of trials
in which the service reports the trap as a document fact.}
\label{fig:propagation}
\end{figure}

\subsection{RQ2: Propagation into LLM Outputs}
\label{sec:rq2}

RQ2 asks whether a trap an extractor emits is actually adopted by a model as a document
fact. We put all 21 mechanisms through eleven interfaces: four native-ingestion APIs
and seven web chatbots. Figure~\ref{fig:propagation} gives every cell.
On our mechanism-balanced set, mean record-level propagation ranges from 0.48 (GLM)
to 0.76 (Kimi) across the APIs and from 0.52 to 0.76 across the web chatbots. Each
mechanism contributes ten records equally, with API outcomes averaged over ten
repetitions, so these rates characterize our evaluation set rather than real-world
attack prevalence.
Across the 840 matched-clean document--API trials, no response surfaces the absent
trap (0/840). Conversely, all four APIs surface the designated target in every
visible-positive trial (24/24 document--API trials). These controls bound the main
condition: the targets are not produced on their matched clean twins, yet are
recoverable through all four APIs when present in ordinary visible content.

\begin{figure}[]
\centering
\includegraphics[width=\columnwidth]{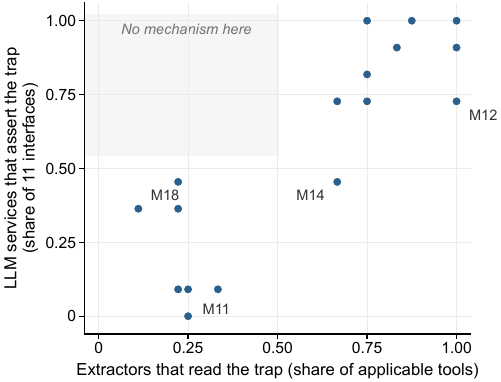}
\caption{Extractor reach and LLM-interface exposure. Each point is one mechanism:
how many applicable panel tools emit its trap, against how many of the eleven LLM
interfaces assert that trap as a document fact.}
\label{fig:ceiling}
\end{figure}

\vspace*{2pt}
\noindent\textbf{Extractor Reach Tracks LLM-Interface Exposure }
Figure~\ref{fig:ceiling} shows a strong observed association between the two layers.
Of the fourteen mechanisms that most applicable panel tools read, every one is asserted
by between five and eleven of the eleven LLM interfaces. Of the seven that few panel tools
read, none reaches more than five interfaces. Mechanisms with broader panel reach thus tend
to appear across more LLM interfaces. This relationship is observational: an LLM service
may use an extractor or configuration outside our panel, and broad panel reach does
not guarantee broad propagation. The table description (M18), for example, is read by
exactly one panel tool but propagates on four interfaces. Overall, twenty of the twenty-one
mechanisms are asserted as fact by at least one interface; the exception is the worksheet
header (M11), which offline tools read but no tested service surfaces as model evidence.

\begin{figure}[t]
\centering
\includegraphics[width=\columnwidth]{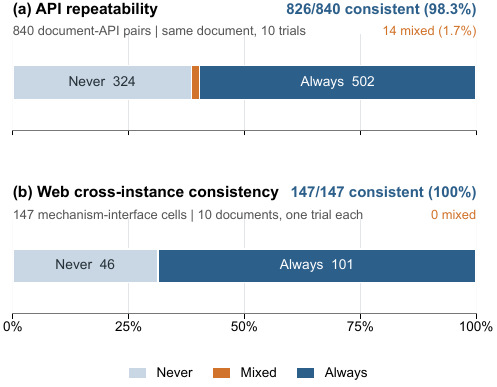}
\caption{Stability under two evaluation designs. API units repeat the same document
ten times; web units contain ten distinct documents tested once each. \emph{Never}
and \emph{Always} are consistent outcomes; \emph{Mixed} contains both verdicts.}
\label{fig:stability}
\end{figure}

\vspace*{2pt}
\noindent\textbf{Propagation Is Stable Across Tested Conditions }
The API and web experiments provide two distinct forms of stability evidence,
summarized in Figure~\ref{fig:stability}. For the APIs, each document is queried ten
times. Of the 840 document--API pairs, 826
(98.3\%) produce the same binary verdict in all ten trials; 502 always surface the
trap, 324 never do, and 14 vary across repetitions. For the web chatbots, each
mechanism--interface cell contains ten distinct documents tested once each. All 147
cells are consistent across those instances: 101 surface the trap for all ten
documents and 46 for none. Thus, the API results measure within-document
repeatability, whereas the web results measure consistency across document instances.

\vspace*{2pt}
\noindent\textbf{Aggregate Web Coverage Is Broader }
Across the two panels in Figure~\ref{fig:propagation}, the seven web interfaces
collectively expose 20 of the 21 mechanisms, while the four APIs collectively expose
18. The difference is M15 and M16, the two Markup-Compatibility branches, which appear
only in the web panel. This
containment holds only for the aggregate unions: it does not imply that a vendor's API
is a lower bound on its web exposure. As RQ3 shows, GPT's two entry paths differ in both
directions, while Qwen's API exposes M21 and its web interface does not.

\begin{findingbox}
\textbf{Finding II.} Broad upstream reach tends to become broad downstream
exposure: LLM interfaces generally preserve, rather than reconcile, the evidence
selected during ingestion. Persistence across API retries and distinct web
submissions points to the pipeline, not generation noise.
\end{findingbox}
\subsection{RQ3: Model or Ingestion Layer?}
\label{sec:rq3}

An LLM interface couples a model with an ingestion path. We ask which of the two more
closely tracks the observed exposure pattern.

\vspace*{2pt}
\noindent\textbf{Same Vendor, Different Entry Points }
Figure~\ref{fig:propagation} shows that one service can behave differently at its two entry
points. GLM's web interface and file-parsing API both advertise GLM~5.2, yet the web
interface exposes five mechanisms (M5, M12, M13, M20, M21) that the API does not, while
the API exposes nothing that the web does not. Holding the advertised model label
constant localizes this difference to the end-to-end entry path, although it does not
identify which backend component differs.
GPT-5.5 also differs across its two entry points, though by less: its web interface exposes
eleven mechanisms and its API fourteen. What stands out on the API is three mechanisms (M5, M12, M14), that it exposes
only rarely, on four to six of a hundred runs rather than all or none.
We re-ran these three and asked the model to return the document text verbatim. The trap
value appeared in the returned text in only a small fraction of runs while the surrounding
text stayed the same across runs. This pattern implicates upstream file handling rather
than ordinary answer stochasticity, although it does not identify the responsible
backend component.

Not every service splits this way. Kimi~K2.6 exposes the same sixteen mechanisms through its
web interface and its API, a result consistent with a shared or behaviorally equivalent
ingestion path. Qwen
differs at a single mechanism (M21), exposed in 20\% of API trials but not through the
web interface; this comparison is weaker than the others, because Qwen's native file-parsing API is
Qwen-doc-turbo while its web chat offers no Qwen-doc-turbo, so the web tests use Qwen-3.5 and
the two ends differ in model as well as path. Taken together, the same vendor's API and web
chat can read the same file differently.

Exposure is far steadier across versions of one model than across entry points. Holding the
OpenAI file-input path fixed, we ran the full corpus of 210 documents ten times each on
\texttt{gpt-5.4} and again on \texttt{gpt-5.5}, 2{,}100 judged runs per model. Not one
mechanism flips. A version apart on the same path, the extraction behaves the
same and so does the attack surface. Across these controls, exposure tracks the tested
ingestion path more closely than model version alone. The clearest case is one where
that path is published, which we turn to next.

\vspace*{2pt}
\noindent\textbf{Three Claude Models, One Skill }
The strongest form of this control is inside a single vendor. Anthropic serves Office
documents to claude application through a document skill, never through its API, which rejects OOXML
at validation; the skill reads XLSX with \texttt{openpyxl}, DOCX with \texttt{pandoc}, and
PPTX with \texttt{markitdown}~\cite{claudeofficeskills}, so the text a model sees is fixed
before it reasons. We ran
three Claude models spanning that vendor's range, Haiku~4.5, Sonnet~5, and Opus~4.8, through
this shared path, and all three expose the same sixteen of the twenty-one mechanisms,
mechanism for mechanism (Table~\ref{tab:claude}). Under this shared path, model choice
does not change the exposed set.

\begin{table}[]
\centering
\footnotesize
\renewcommand{\arraystretch}{1.15}
\caption{Three Claude models of different capability, one ingestion path. All three read
Office through the same claude.ai document skill (\texttt{openpyxl}/\texttt{pandoc}/%
\texttt{markitdown}).}
\label{tab:claude}
\begin{tabular}{l r}
\toprule
\textbf{Model} & \textbf{Mechanisms exposed} \\
\midrule
Haiku 4.5   & 16/21 \\
Sonnet 5  & 16/21 \\
Opus 4.8  & 16/21 \\
\midrule
Same exposed set & M1--M10, M12--M13, M17--M20 \\
\bottomrule
\end{tabular}
\end{table}

\vspace*{2pt}
\noindent\textbf{Version and Invocation Change What a Tool Reads }
Claude reads DOCX through \texttt{pandoc}, and so does our panel, yet on one mechanism the two
disagree. Our \texttt{pandoc} reads the off-page text box (M14) on every sample
(Table~\ref{tab:reach}), while Claude Sonnet~5 never exposes it
(Figure~\ref{fig:propagation}). Claude's published skill confirms the library, but
its behavior differs from our pinned version. Reading M14
changed between the two \texttt{pandoc} releases we ran, 3.8.3 missing it on every sample and
3.10 reading it on every one, so our 3.10 panel surfaces M14 while Claude's pattern is
consistent with an older build. Two more cases have the same shape. A hyperlink target (M19) is dropped by
\texttt{pandoc -t plain} but emitted by \texttt{-t markdown}, the format an LLM pipeline
actually wants. A merged-cell value (M20) is masked by \texttt{openpyxl}'s default load yet
surfaced in its streaming \texttt{read\_only} mode, the one used to dump large sheets.
Version, output format, and read mode change which document evidence reaches the
model, so knowing that a service
uses \texttt{pandoc} or \texttt{openpyxl} says little until we also know the version it pins
and the arguments it passes.

\begin{figure}[t]
\centering
\includegraphics[width=\columnwidth]{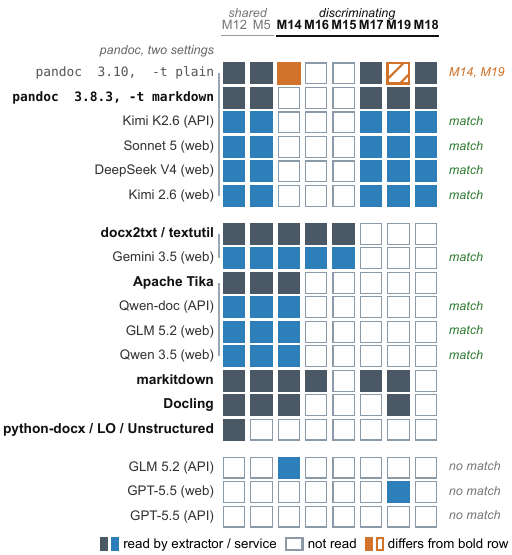}
\caption{Behavior-level matching between LLM interfaces and candidate DOCX extraction
configurations. Dark rows denote tested configurations; blue rows share their
eight-probe vectors. A match does not identify the backend.}
\label{fig:fingerprint}
\end{figure}

\vspace*{2pt}
\noindent\textbf{Behavior-Level Matching of Closed Ingestion Paths }
A mechanism can be read differently not only by different extraction tools but by different versions
and arguments of one tool, so the set of mechanisms a service surfaces carries information about
how it reads documents. We therefore ask whether that pattern can match a closed service
to tested candidate configurations, including versions and invocation modes. The pattern
is usable because each tool's per-mechanism outcome is deterministic, making it a stable
behavioral signature over our probes. Claude's published skill, for instance, names three
libraries but not the version or read mode each runs at, and a fully closed service names
none, so there is useful behavior to compare.
We run this test on DOCX, because that is where the extractors diverge. In
Table~\ref{tab:reach} the eight spreadsheet tools collapse to six signatures, three of them
identical, so an XLSX pattern separates almost nothing. The Word tools instead each read a
different subset: of the eight DOCX mechanisms, two are read by nearly every tool and tell us
nothing, while the other six split the tools apart, and one of those, the table description
(M18), is read by \texttt{pandoc} alone. A DOCX pattern is therefore a useful
behavioral signature
(Figure~\ref{fig:fingerprint}).

Matched against these signatures, 8 of the 11 interfaces match at least
one tested candidate configuration over the eight probes. Four share a vector with
\texttt{pandoc}: Kimi K2.6 (API), Sonnet~5 (web), DeepSeek V4 (web), and Kimi K2.6
(web). Gemini 3.5 (web) matches \texttt{docx2txt} or \texttt{textutil}; Qwen-doc
(API), GLM 5.2 (web), and Qwen 3.5 (web) match Apache Tika. The four
\texttt{pandoc}-like interfaces can be matched to a more specific tested configuration:
their vector differs from our pinned \texttt{pandoc} at two mechanisms, while
\texttt{pandoc} 3.8.3 with Markdown output reproduces it on all eight. Markdown carries
the hyperlink target (M19), and only the newer version reads the off-page text box
(M14). This establishes behavioral equivalence over our probes, not backend identity.

Claude provides the only external check: its published skill names \texttt{pandoc},
and its observed vector matches a tested \texttt{pandoc} configuration. This validates
the library-level match but not the exact version or invocation, so we claim an
existence proof of behavior-level candidate matching rather than a fingerprinting
benchmark. The more mechanisms a panel distinguishes, the more tightly it can narrow
the configurations consistent with a service's behavior. An attacker who cannot see a
service's stack can use a fixed probe set to narrow those candidates and prioritize
attacks known to affect them without learning the backend's identity.

\begin{findingbox}
\textbf{Finding III.} Model identity and extractor name are both too coarse:
evidence-fork exposure depends on the full ingestion configuration, including
version, output mode, and arguments. The resulting stable exposure vector lets
us use evidence forks to narrow a closed service to behaviorally equivalent
candidate configurations without identifying its backend.
\end{findingbox}

\subsection{RQ4: How Widely Do Affected Ingestion Paths Appear in Open-Source Projects?}
\label{sec:rq4}

We answer this with a source-level survey, not deployment telemetry or end-to-end
measurements. At one pinned commit per repository (July~2026), we inventory the default
OOXML ingestion paths of sixteen popular open-source agent and RAG projects and
follow each import chain to the \emph{leaf} library that turns OOXML into text. These
mappings identify reachable extractor families, not propagation through the complete
application (Table~\ref{tab:frameworks}).
We stop at a library only when it does the extraction itself. markitdown, Unstructured, and
Apache Tika are recorded as they are called, because each rewrites the document into its own
normalized text with its own reach. A thin pass-through is instead resolved to what it
actually loads: the six applications that read spreadsheets through \texttt{pandas}
(LangChain, LlamaIndex, Haystack, Dify, Kotaemon, Verba) are listed as \texttt{openpyxl}, the
engine \texttt{pandas} loads by default, since none selects the alternative calamine engine
and \texttt{pandas} contributes no extraction of its own.

The sixteen projects concentrate on a few leaf extractor families. Their default paths
therefore map to families with the reach profiles measured in RQ1, although application
versions, invocation arguments, and downstream processing may change the resulting
exposure. markitdown has the widest panel reach, 17 of the 21 mechanisms, and four
projects route through it, including AutoGen, GraphRAG, and PrivateGPT. Separately,
txtai is the only surveyed project whose default path reaches Apache Tika (16/21),
the only panel tool that surfaces the phonetic guide and worksheet header.

The source mappings preserve RQ1's format asymmetry: Word paths span families with
widely different measured reach, whereas the spreadsheet families all have broad panel
reach. Switching among the surveyed spreadsheet defaults is therefore unlikely to
eliminate potential exposure, although exact application behavior still depends on
versions, arguments, and downstream processing.

\begin{table}[]
\centering
\footnotesize
\setlength{\tabcolsep}{7pt}
\renewcommand{\arraystretch}{1.15}
\caption{The leaf extractor(s) reached by each surveyed project's default OOXML
ingestion path, as resolved from source code. GitHub stars were measured on
July~26, 2026.}
\label{tab:frameworks}
\begin{tabular}{l l r}
\toprule
\textbf{Application} & \textbf{Leaf extractor(s)} & \textbf{Stars} \\
\midrule
AutoGen~\cite{autogen} & markitdown & 60.0K \\
GraphRAG~\cite{graphrag} & markitdown & 34.9K \\
PrivateGPT~\cite{privategpt} & markitdown & 57.4K \\
Onyx~\cite{onyx} & markitdown, openpyxl & 31.2K \\
LangChain~\cite{langchain} & docx2txt, openpyxl, python-pptx & 142.6K \\
LlamaIndex~\cite{llamaindex} & docx2txt, openpyxl, python-pptx & 51.1K \\
Quivr~\cite{quivr} & docx2txt, Unstructured & 39.4K \\
Embedchain~\cite{embedchain} & docx2txt, Unstructured & 61.8K \\
Haystack~\cite{haystack} & python-docx, openpyxl, python-pptx & 26.0K \\
RAGFlow~\cite{ragflow} & python-docx, openpyxl, python-pptx & 86.1K \\
R2R~\cite{r2r} & python-docx, openpyxl, python-pptx & 7.9K \\
Dify~\cite{dify} & python-docx, openpyxl, Unstructured & 150.3K \\
Verba~\cite{verba} & python-docx, openpyxl & 7.7K \\
CrewAI~\cite{crewai} & python-docx & 56.2K \\
Kotaemon~\cite{kotaemon} & Unstructured, openpyxl & 25.6K \\
txtai~\cite{txtai} & Apache Tika & 12.8K \\
\bottomrule
\end{tabular}
\end{table}

\begin{findingbox}
\textbf{Finding IV.} Project diversity masks dependency concentration:
default OOXML paths converge on a small set of shared extractor families.
This concentration has format-dependent consequences: changing the Word
loader can substantially alter potential exposure, whereas the surveyed
spreadsheet paths all resolve to families with broad panel reach.
\end{findingbox}

%% file: figs/tab_reach.tex
\begin{table*}[t]
\centering
\small
\renewcommand{\arraystretch}{1.1}
\newcommand{\ee}{\CIRCLE}
\newcommand{\oo}{\Circle}
\newcommand{\rot}[1]{\rotatebox[origin=l]{65}{#1}}
\caption{Extractor reach. \ee~= the tool emits the trap; \oo~= it handles the format but
its output stays benign; blank~= it does not handle the format. Denominators are
format-aware: \textbf{Reach} counts tools per mechanism (8 XLSX, 9 DOCX, 6 PPTX), and the
bottom row counts mechanisms per tool.}
\label{tab:reach}
\resizebox{1\textwidth}{!}{
\begin{tabular}{l l ccc cccc c ccccc | c}
\toprule
Mechanism & Format & \rot{openpyxl} & \rot{calamine} & \rot{xlsx2csv} & \rot{python-docx} & \rot{docx2txt} & \rot{pandoc} & \rot{textutil} & \rot{python-pptx} & \rot{markitdown} & \rot{Apache Tika} & \rot{LibreOffice} & \rot{Unstructured} & \rot{Docling} & Reach \\
\midrule
M1   & xlsx & \ee  & \ee  & \ee  &      &      &      &      &      & \ee  & \ee  & \ee  & \ee  & \ee  & 8/8 \\
M2   & xlsx & \ee  & \ee  & \ee  &      &      &      &      &      & \ee  & \ee  & \ee  & \ee  & \ee  & 8/8 \\
M3   & xlsx & \ee  & \ee  & \ee  &      &      &      &      &      & \ee  & \ee  & \ee  & \ee  & \ee  & 8/8 \\
M4   & xlsx & \ee  & \ee  & \ee  &      &      &      &      &      & \ee  & \ee  & \ee  & \ee  & \ee  & 8/8 \\
M5   & docx &      &      &      & \oo  & \ee  & \ee  & \ee  &      & \ee  & \ee  & \oo  & \oo  & \ee  & 6/9 \\
M6   & xlsx & \ee  & \ee  & \ee  &      &      &      &      &      & \ee  & \ee  & \oo  & \ee  & \oo  & 6/8 \\
M7   & xlsx & \ee  & \ee  & \ee  &      &      &      &      &      & \ee  & \ee  & \oo  & \ee  & \oo  & 6/8 \\
M8   & xlsx & \ee  & \ee  & \oo  &      &      &      &      &      & \ee  & \ee  & \ee  & \ee  & \ee  & 7/8 \\
M9   & xlsx & \ee  & \ee  & \oo  &      &      &      &      &      & \ee  & \ee  & \ee  & \ee  & \ee  & 7/8 \\
M10  & xlsx & \ee  & \ee  & \ee  &      &      &      &      &      & \ee  & \ee  & \ee  & \ee  & \ee  & 8/8 \\
M11  & xlsx & \ee  & \oo  & \oo  &      &      &      &      &      & \oo  & \ee  & \oo  & \oo  & \oo  & 2/8 \\
M12  & docx &      &      &      & \ee  & \ee  & \ee  & \ee  &      & \ee  & \ee  & \ee  & \ee  & \ee  & 9/9 \\
M13  & pptx &      &      &      &      &      &      &      & \ee  & \ee  & \ee  & \oo  & \ee  & \ee  & 5/6 \\
M14  & docx &      &      &      & \oo  & \ee  & \ee  & \ee  &      & \ee  & \ee  & \oo  & \oo  & \ee  & 6/9 \\
M15  & docx &      &      &      & \oo  & \ee  & \oo  & \ee  &      & \oo  & \oo  & \oo  & \oo  & \oo  & 2/9 \\
M16  & docx &      &      &      & \oo  & \ee  & \oo  & \ee  &      & \ee  & \oo  & \oo  & \oo  & \oo  & 3/9 \\
M17  & docx &      &      &      & \oo  & \oo  & \ee  & \oo  &      & \ee  & \oo  & \oo  & \oo  & \oo  & 2/9 \\
M18  & docx &      &      &      & \oo  & \oo  & \ee  & \oo  &      & \oo  & \oo  & \oo  & \oo  & \oo  & 1/9 \\
M19  & docx &      &      &      & \oo  & \oo  & \oo  & \oo  &      & \ee  & \oo  & \oo  & \oo  & \ee  & 2/9 \\
M20  & xlsx & \oo  & \ee  & \ee  &      &      &      &      &      & \ee  & \ee  & \ee  & \ee  & \oo  & 6/8 \\
M21  & xlsx & \oo  & \oo  & \ee  &      &      &      &      &      & \oo  & \ee  & \oo  & \oo  & \oo  & 2/8 \\
\midrule
\multicolumn{2}{l}{Exposed} & 10/12 & 10/12 & 9/12 & 1/8 & 5/8 & 5/8 & 5/8 & 1/1 & 17/21 & 16/21 & 9/21 & 12/21 & 12/21 & \\
\bottomrule
\end{tabular}
}
\end{table*}

%% file: pages/7.defense.tex
\subsection{RQ5: How Common Are These Mechanisms in Ordinary Documents?}
\label{sec:rq5}

A defense that only strips invisible text misses most of these mechanisms, because most leave
both the shown value and the extracted value in the file, while the ingestion path exposes
only one to the model. What matters is not whether text is hidden but how often each construction
appears in ordinary documents: rare constructions can support selective triage, while
common ones cannot be distinguished by presence alone. RQ5 measures this base rate.
We write one static rule per mechanism that flags its construction, and run the probe over
fixed-seed random samples of 2{,}000 XLSX files from
Sheetpedia~\cite{sheetpedia} and 2{,}000 DOCX files from
DocxCorp~\cite{docxcorp}. Both are public datasets with stable dataset records and
downloadable manifests. For PowerPoint, we retain a separate set of 263 public PPTX
files collected from national-government portals and research
repositories~\cite{datagovuk,opencanada,datagovau,harvarddataverse,zenodo,figshare}.
This smaller PPTX set is heterogeneous and serves only as the denominator for M13.
We validate every retained file as an OOXML package and remove exact duplicates by
SHA-256. Each rule fires on all constructed evidence-fork instances,
so it matches the construction faithfully; on ordinary files the same rule
measures how often that construction occurs on its own, not whether it is malicious. Every
construction has legitimate uses, a hidden sheet holding a lookup table, an image carrying
alt-text, a merged title cell, so the result is a base rate.

\begin{figure}[t]
\centering
\includegraphics[width=\columnwidth]{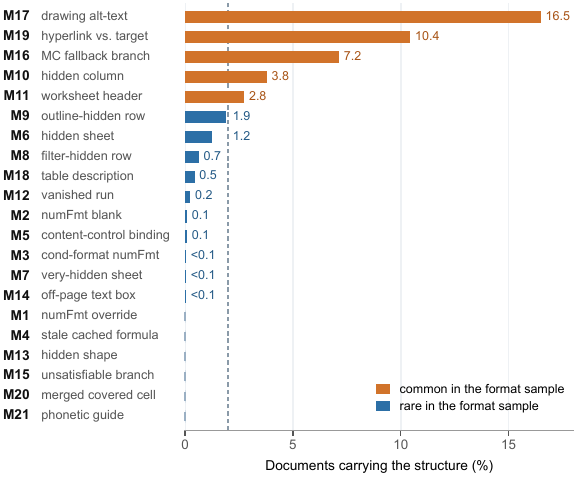}
\caption{Share of documents carrying each mechanism's structural signature, measured
against its own format sample: 2{,}000 Sheetpedia XLSX, 2{,}000 DocxCorp DOCX, or
263 public PPTX files.}
\label{fig:defense}
\end{figure}

The base rates are sharply bimodal (Figure~\ref{fig:defense}). Six mechanisms were not
observed in any file in their corresponding format samples, and ten more appeared in
under 2\%. The remaining five range
from 2.75\% to 16.5\%: worksheet headers, hidden columns, compatibility fallbacks,
labeled hyperlinks, and image alt-text. Across the three format-specific samples,
748 of 4{,}263 documents (17.5\%) carry at least one signature. Because the format
sample sizes are chosen independently, this aggregate is descriptive rather than a
population estimate.

The split says what a presence scan can and cannot do. The sixteen rare signatures
provide selective triage signals in these samples; this covers every number-format
override, stale cache, data binding, merged-cell value, and hidden sheet, row, or shape.
The five common signatures mix metadata and non-body channels with routine spreadsheet
state, and occur in 2.75\% to 16.5\% of their format samples. A presence flag is therefore
not actionable for them. For these common constructions, malicious and benign documents can
use the same structural feature; their difference lies in the content and its intended
role, which presence alone does not capture. Telling attack from benign use there
needs the content itself, or provenance that marks text
drawn from a channel the reader never sees so a downstream model can discount it rather than
accept it as body content.

\begin{findingbox}
\textbf{Finding V.} Benign base rates split the mechanisms into two regimes for presence-only screening: structural presence is a usable triage signal for the rare constructions but cannot separate a planted fork from ordinary use on the common channels. Discriminating the latter requires content and role, not presence alone.
\end{findingbox}

%% file: pages/7.discussion.tex
\section{Discussion and Limitations}
\label{sec:discussion}

\noindent\textbf{View Construction Is a Security Policy }
An OOXML loader is usually treated as a preprocessing utility, but it decides
which document view becomes model evidence. RQ3 shows that this decision belongs
to the full ingestion configuration: entry point, extractor version, output mode,
and arguments can all change exposure, while models sharing one path can retain
the same exposed set. RQ4 shows why this matters beyond one service: otherwise
diverse projects repeatedly route OOXML through a small set of extractor families.
Loader selection and upgrades should therefore be treated as security-relevant
policy changes rather than interchangeable implementation details. The stable
vectors observed under API repetition and across web instances also suggest a
practical audit primitive: a fixed evidence-fork suite can regression-test a
pipeline's evidence-selection behavior and detect drift without backend access.

\vspace*{2pt}
\noindent\textbf{Coverage Is Not Semantic Fidelity }
RQ1 shows that recovering more strings can widen the set of semantic roles
flattened into model evidence. Narrow extraction is not a general remedy:
omitting document semantics does not restore fidelity to the Office view. Text
coverage and semantic fidelity are therefore distinct objectives.
RQ5 shows why
defense cannot stop at structural presence: rare constructions can support
selective alerts, while common channels also serve routine document functions.
A promising direction for future work is a two-part defense: cheap presence
checks for rare constructions, paired with role-aware consistency checks for
common ones. Such a scheme would likely need emitted evidence to retain its
source and semantic role, and the ingestion contract to make representation
selection explicit.  We leave building and evaluating it to future work.

\vspace*{2pt}
\noindent\textbf{Limitations }
Because each mechanism contributes equally, aggregate propagation rates
characterize our mechanism-balanced evaluation set rather than the distribution
of documents in deployment. We therefore interpret them together with
per-mechanism outcomes and interface coverage, rather than as population-level
risk estimates.
Results for tools, services, and project dependencies reflect the pinned or
dated configurations evaluated in this study. Implementations may evolve and
expose different subsets, so product-level mappings are snapshots rather than
permanent rankings. The evidence-fork catalog and analysis method do not depend
on any one vendor configuration.

%% file: pages/8.conclusion.tex
\section{Conclusion}
\label{sec:conclusion}

OOXML-to-LLM pipelines can construct model evidence that diverges from the Office
view presented for review. We systematized this failure as evidence forks
and confirmed 21 specification-grounded mechanisms across Excel, Word, and
PowerPoint. Cross-stack measurements show that exposure follows the ingestion
configuration, not model identity alone, and that shared extractor families recur
across open-source projects. Secure ingestion therefore needs an explicit contract
that defines the intended view, preserves evidence provenance and semantic roles,
and surfaces rather than silently flattens disagreements. Evidence-fork probes
provide a practical way to test that contract as loaders evolve.

%% file: pages/9.ethics.tex
\section*{Ethical Considerations}
\label{sec:ethics}

\noindent\textbf{Stakeholders and Impacts }
The stakeholders are users and organizations that rely on OOXML-to-LLM workflows;
LLM providers and extractor maintainers; other users of the measured services; the
research team; and the broader security community. The principal benefit is to expose
an ingestion risk that is otherwise invisible to document reviewers and to support
defensive testing. Potential harms include the provider cost of live measurements,
the risk of over-attributing behavior to a particular service, and dual-use knowledge
that could help create misleading documents. We considered beneficence, respect for
persons, and respect for law and the public interest in designing the study and the
mitigations below.

\vspace*{2pt}
\noindent\textbf{Live-System Testing }
Offline extractors cannot establish how proprietary end-to-end ingestion paths handle
an uploaded document, so the RQ2 and RQ3 measurements required limited testing of live
services. We used the authors' own accounts and only the API and web-upload
functionality available to those accounts. All documents contained controlled,
non-harmful factual substitutions or summary statements, and all prompts requested
ordinary factual question answering or summarization. We included no attack
instructions, prompt injections, jailbreaks, malware, macros, or requests for harmful
content. We did not attempt penetration, account compromise, bypass of access
controls, data exfiltration, service disruption, or access to other users' or
confidential data. Testing was limited to the pre-specified trial matrix in
\S\ref{sec:setup}; we did not stress-test the services or intentionally alter shared
state. The remaining provider impact was the ordinary computation required to process
our uploads and queries.

\vspace*{2pt}
\noindent\textbf{Human Subjects and Data }
We recruited no external participants and collected no data from service users; human
review was performed only by the authors on Office renderings and model responses. The
realistic fixtures use excerpts from the public TAT-QA research
dataset~\cite{zhu2021tatqa} and plant synthetic traps solely for controlled
evaluation. The RQ5 study scans public datasets and documents locally and reports
structural signatures and aggregate rates. We did not collect private user documents
or intentionally extract or publish personal information that might appear in those
public files. Because the tasks were benign financial question answering and
summarization, manual review did not expose the research team to harmful or disturbing
content.

\vspace*{2pt}
\noindent\textbf{Provider Feedback }
We do not frame the measured behavior as a conventional software vulnerability or
parser bug: the documents are specification-valid, and the risk arises across OOXML
features and ingestion decisions rather than from one faulty component. Nevertheless,
before submission we submitted feedback to several tested LLM providers through their
available product-feedback or security channels so that they could evaluate their
ingestion behavior. These notifications are not used as evidence for our claims, and
any acknowledgment, scope decision, or lack of response does not validate or refute
the measurements.

\vspace*{2pt}
\noindent\textbf{Dual Use and Decision to Publish }
The catalog and fixture generator are functional dual-use artifacts: they can support
regression testing, but they could also lower the effort needed to craft a misleading
Office document. We mitigate this risk by using only benign factual traps, excluding
harmful instructions and executable payloads, reporting provenance-aware defenses,
and notifying providers before publication. Publication cannot eliminate misuse risk.
We nevertheless decided to conduct and publish the study because closed ingestion
behavior cannot be evaluated from offline components alone, the underlying OOXML
features are already publicly documented, and a reproducible benchmark enables
providers, maintainers, and users to identify and reduce the risk. We judge these
defensive benefits to outweigh the residual live-service and dual-use costs under the
safeguards above.